\documentclass[11pt]{article}
\usepackage{jeosman}
\usepackage{amsmath}
\usepackage{amssymb}
\usepackage{textcomp}

\def\##1{{\underline #1}}
\def\~#1{{\underline {\mathcal#1}}}
\def\+#1{{{\mathcal #1}}}
\def\=#1{\underline{\underline #1}}

\def\.{\mbox{ \tiny{$^\bullet$} }}

\def\eps{\epsilon}
\def\epso{\epsilon_o}
\def\epsa{\epsilon_a}
\def\epsb{\epsilon_b}
\def\epsc{\epsilon_c}
\def\epsd{\epsilon_d}

\def\ux{\#u_x}
\def\uy{\#u_y}
\def\uz{\#u_z}

\def\muo{\mu_o}
\def\lambdao{\lambda_o}
\def\etao{\eta_o}
\def\ko{k_o}

\def\cpz{\cos\zeta}
\def\spz{\sin\zeta}
\def\cpzsq{\cos^2\zeta}
\def\spzsq{\sin^2\zeta}

\def\le{\left(}
\def\ri{\right)}
\def\les{\left[}
\def\ris{\right]}

\begin{document}


\title{Dyakonov--Tamm wave at the planar interface of a chiral sculptured thin film  and an isotropic dielectric material}

\author{Akhlesh Lakhtakia}
\address{CATMAS --- Computational \& Theoretical Materials Sciences Group, Department of Engineering Science and Mechanics, Pennsylvania State University, University Park, PA 16802, USA}
\email{akhlesh@psu.edu}

\vskip 18pt

\author{John A. Polo, Jr.}
\address{Department of Physics and Technology,
Edinboro University of Pennsylvania,
Edinboro, PA 16444, USA}
\email{polo@edinboro.edu}

\begin{abstract}Surface waves, named here as Dyakonov--Tamm waves, can exist at the planar interface of an isotropic dielectric material
and a chiral sculptured thin film (STF). Due to the periodic nonhomogeneity of a
chiral STF, the range of the refractive index of the isotropic material is smaller
but the range of the propagation direction in the interface plane is much larger, in comparison
to those for the existence of Dyakonov waves at the planar interface of an isotropic dielectric material
and a columnar thin film.
\vskip 6pt

  \end{abstract}


\section{INTRODUCTION}
\label{sect:intro}  

Less than two decades ago, Dyakonov \cite{Dyakonov1988} theoretically predicted the propagation
of a surface wave at the planar interface of an isotropic dielectric material
and a positively uniaxial dielectric material with its optic axis wholly parallel to the
interface plane.  If $\psi$ indicates  the angle between the optic axis and the
direction of surface--wave propagation, and $n_s$ is the refractive index of the isotropic
dielectric material, then the \emph{Dyakonov wave} exists for rather narrow ranges
of $\psi$ and $n_s$. The consequent significance of Dyakonov waves for optical sensing and waveguiding was recognized thereafter \cite{Torner1993,Torner1995}.
Since then, the concept of the Dyakonov wave has been extended to the planar interfaces
of isotropic and biaxial dielectric materials \cite{Walker1998}. The possibility of the anisotropic material being artificially engineered, either as a photonic crystal
with a short period in comparison to the wavelength \cite{Artigas2005} or as
a columnar thin film (CTF) \cite{PNL2007},
 has also emerged. Let us note here that  the Dyakonov wave still remains
to be experimentally observed, in part due to the narrow range of $\psi$ for its
existence \cite{Artigas2005}.

The anisotropic material is taken to be homogeneous in all of the foregoing and other reports on the Dyakonov wave.   What if the anisotropic material were to be chosen
as periodically nonhomogeneous
in a direction normal to the bimaterial interface? This question initiated a research
project, the first results of which are being communicated here. Being a natural
extension of a CTF, a chiral
sculptured thin film (STF) was chosen as the periodically nonhomogeneous
anisotropic material \cite{STFbook}.

\begin{figure}
    \begin{center}
    \begin{tabular}{c}
    \includegraphics[height=5cm]{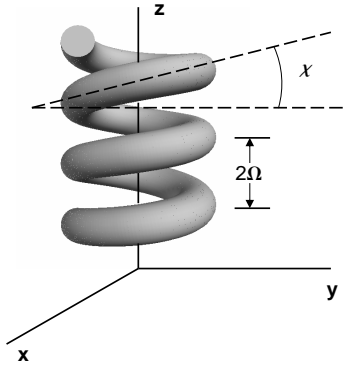}
    \end{tabular}
    \end{center}
    \caption{ \label{Fig:chiSTF} Geometry of a structurally right--handed helix.}
 \end{figure}

A chiral STF  is made by directing a vapor flux in vacuum at an oblique angle  onto a rotating
substrate.  Under suitable conditions, an assembly of parallel nanohelixes of the evaporated species
forms, with the helical axes perpendicular to the substrate.  An example of a single nanohelix is
illustrated in Figure~\ref{Fig:chiSTF}.  By adjusting deposition parameters, both the pitch   $2\Omega$
and the angle of inclination $\chi\in(0,\pi/2]$ can be controlled.  Each nanohelix, composed of
multimolecular clusters  with $\sim3$~nm diameter, is effectively a continuously bent column of $\sim
100$-nm  cross--sectional diameter. Therefore, at visible frequencies and lower, a chiral STF may be
regarded as a linear, locally orthorhombic, unidirectionally nonhomogeneous continuum whose relative
permittivity dyadic is akin to that of chiral smectic liquid crystals \cite{deG}.

In formulating the surface--wave--propagation
problem on the planar interface of an isotropic, homogeneous, dielectric material and a chiral STF, we
adopted a methodology originally developed by Tamm in 1932 for a realistic Kronig--Penney model. Instead
of assuming the solid to occupy the entire space,
 as is commonplace in solid--state physics \cite{Kittel},
 Tamm assumed the solid to occupy only a half--space. The incorporation of the
 oft--neglected surface led to the emergence of electronic states localized
 to the surface. \emph{Tamm states} were experimentally observed in 1990 on the surfaces
 of superlattices  \cite{Ohno1990}, and their optical analogs
 for superlattices of isotropic materials are being investigated these days \cite{Martorell2006,Namdar2006}.

Given the braiding of Dyakonov waves and Tamm states in this communication,
we decided to name the surface wave at the planar interface
of an isotropic, homogeneous, dielectric material and a chiral STF as the \emph{Dyakonov--Tamm wave.} Section~\ref{form} presents  the boundary--value problem and the dispersion equation for the Dyakonov--Tamm wave. Section~\ref{res} contains
numerical results when the chiral STF is chosen to be made of titanium oxide
\cite{PNL2007,HWH1998}. An $\exp(-i\omega t)$ time--dependence is implicit, with $\omega$
denoting the angular frequency. The free--space wavenumber, the
free--space wavelength, and the intrinsic impedance of free space are denoted by $\ko=\omega\sqrt{\epso\muo}$,
$\lambdao=2\pi/\ko$, and
$\etao=\sqrt{\muo/\epso}$, respectively, with $\muo$ and $\epso$ being  the permeability and permittivity of
free space. Vectors are underlined, dyadics underlined twice;
column vectors are underlined and enclosed within square brackets, while
matrixes are underlined twice and similarly bracketed. Cartesian unit vectors are
identified as $\ux$, $\uy$ and $\uz$.

\section{FORMULATION}\label{form}

\subsection{Geometry and permittivity}

Let the half--space $z\leq0$ be occupied by an isotropic, homogeneous, nondissipative, dielectric
material of refractive index $n_s$. The region $z\geq0$ is occupied by a chiral STF with
unidirectionally nonhomogeneous permittivity dyadic given by \cite{STFbook}
\begin{equation}
\=\epsilon(z)= \epso \, \=S_z(z)\cdot\=S_y(\chi)\cdot\=\epsilon_{\rm
ref}\cdot\=S_y^T(\chi)\cdot\=S_z^T(z)
\, , \quad z\geq 0 \,,
\end{equation}
where  the reference relative permittivity dyadic
\begin{equation}
\=\epsilon_{ref} = \epsa  \, \uz\uz  + \epsb \, \ux\ux
+ \epsc \, \uy\uy\,
\end{equation}
indicates the locally orthorhombic symmetry of the chiral STF.
The
dyadic function
\begin{eqnarray}
\nonumber
\=S_z(z)&=&\cos{\le \frac{\pi z}{\Omega} +\psi \ri}\le \ux\ux + \uy\uy \ri \\
&&+\, h\,\sin{\le  \frac{\pi z}{\Omega} +\psi \ri}\le \uy\ux - \ux\uy \ri+\uz\uz \,\label{SzDyadic}
\end{eqnarray}
contains $2\Omega$ as the structural
period, $\psi$ as an angular offset, and $h=\pm 1$ as the handedness parameter.
The tilt dyadic
\begin{eqnarray}
\nonumber
\=S_y(\chi)&=&\le \ux\ux + \uz\uz \ri \cos{\chi}
\\ &&+\le \uz\ux -
\ux\uz \ri \sin{\chi}+\uy\uy \,
\end{eqnarray}
involves the  angle of inclination $\chi$.   The superscript $^T$ denotes the transpose.

Without loss of generality, we take the Dyakonov--Tamm wave to propagate parallel to the $x$ axis in the plane $z=0$.
There is no dependence on the $y$ coordinate, whereas the Dyakonov--Tamm wave must
attenuate as $z\to\pm\infty$.

\subsection{Field representations}
In the region $z \leq 0$, the wave vector may be written as
\begin{equation}
\#k_s=\kappa\, \ux -\alpha_s\,\uz\,,
\end{equation}
where
\begin{equation}
\kappa^2+\alpha_s^2=\ko^2\, n_s^2\,,
\end{equation}
 $\kappa$ is positive and real--valued for unattenuated
propagation along the $x$ axis, and
 ${\rm Im}\left[\alpha_s\right]>0$ for attenuation as $z\to-\infty$.
Accordingly, the field phasors in the region $z\leq 0$ may be written as
\begin{equation}
\#E(\#r)= \left[A_{1}\,\uy + A_{2}\left( \frac{\alpha_s}{\ko}\,\ux +\frac{\kappa}{\ko}\, \uz\right)\right]
\exp(i\#k_s\cdot\#r)\,,\quad z \leq 0\,,
\label{eqn:Esub}
\end{equation}
and
\begin{equation}
\#H(\#r)=\etao^{-1}\left[ A_{1} \left(\frac{\alpha_s}{\ko}\,\ux+\frac{\kappa}{\ko}\,\uz\right) -
A_{2}\,n_s^2\,\uy\right] \exp(i\#k_s\cdot\#r)\,,\quad z \leq 0\,,\label{eqn:Hsub}
\end{equation}
where $A_{1}$ and $A_{2}$ are unknown scalars.

 The field representation in the region $z\geq0$ is more complicated. It is best to
write
\begin{equation}
\left.\begin{array}{l}
\#E(\#r)=\#e(z)\,\exp(i\kappa x)\\
\#H(\#r)=\#h(z)\,\exp(i\kappa x)
\end{array}\right\}\,,\qquad
z\geq0\,.
\end{equation}
and create the column vector
\begin{equation}
\left[\#f(z)\right]= \left[e_x(z)\quad e_y(z) \quad h_x(z)\quad h_y(z)\right]^T\,.
\end{equation}
This column vector satisfies the matrix differential equation \cite{STFbook}
\begin{equation}
\label{MODE}
\frac{d}{dz}\left[\#f(z)\right]=i \left[\=P(\frac{\pi z}{\Omega} +\psi,\kappa)\right]\cdot\left[\#f(z)\right]\,,
\quad z>0\,,
\end{equation}
where the 4$\times$4 matrix
 \begin{eqnarray}
\nonumber
&&
[\=P(\zeta,\kappa)]=
\\[5pt]
\nonumber
&&\omega\,\les\begin{array}{cccc}
0 & 0 & 0 & \muo \\[4pt]
0 & 0 & -\muo & 0 \\[4pt]
h\,\epso\left(\epsc-\epsd\right)\cpz\,\spz & -\epso\left(\epsc\cpzsq
+\epsd\spzsq\right) & 0 & 0\\[4pt]
\epso\left(\epsc\spzsq +\epsd\cpzsq\right) & -h\,\epso\left(\epsc-
\epsd\right)\cpz\,\spz & 0 & 0
\end{array}\ris
\\[5pt]
\nonumber
&& \qquad
+\,\kappa\,\frac{\epsd\,\left(\epsa-\epsb\right)}{\epsa\,\epsb}\,
\sin\chi\cos\chi\,
\les\begin{array}{cccc}
\cpz  & h\,\spz  & 0 & 0\\[4pt]
0 & 0 & 0 & 0\\[4pt]
0 & 0 & 0& -h\,\spz \\[4pt]
0 & 0 & 0 & \cpz
\end{array}\ris
\\[5pt]
&& \qquad
+
\les\begin{array}{cccc}
0 & 0 & 0 & -\,\frac{\kappa^2}{\omega\epso}\,\frac{\epsd}{\epsa\,\epsb}\\[4pt]
0 & 0 & 0 & 0 \\[4pt]
0 & \frac{\kappa^2}{\omega\muo}& 0 & 0\\[4pt]
0 & 0 & 0 & 0
\end{array}\ris\,
\label{Pdef}
\end{eqnarray}
and
\begin{equation}
\epsd=\frac{\epsa\epsb}{\epsa\,\cos^2\chi+\epsb\,\sin^2\chi}\,.
\end{equation}

Two independent techniques \cite{SchHer,PL04} exist to solve (\ref{MODE}),
which may be
harnessed to determine the matrix $[\=N]$ that appears in the relation
\begin{equation}
[\#f(2\Omega)]=[\=N]\cdot[\#f(0+)]
\end{equation}
to characterize the optical response of one period of the chiral STF. By virtue of the Floquet--Lyapunov theorem
\cite{YS75}, we can define a matrix $[\=Q]$ such that
\begin{equation}
[\=N] = \exp\left\{i2\Omega[\=Q]\right\}\,.
\end{equation}
Both $[\=N]$ and $[\=Q]$ share the same eigenvectors, and their eigenvalues
are also related. Let $[\#t]^{(n)}$, $(n=1,2,3,4)$, be the eigenvector corresponding
to the  $n$th eigenvalue $\sigma_n$ of $[\=N]$; then, the corresponding eigenvalue
$\alpha_n$ of $[\=Q]$
is given by
\begin{equation}
\alpha_n = -i\frac{\ln \sigma_n}{2\Omega}\,.
\end{equation}

\subsection{Dispersion equation for Dyakonov--Tamm wave}
For the Dyakonov--Tamm wave to propagate along the $x$ axis, we must ensure that ${\rm Im}[{\alpha_{1,2}}]>0$, and set
\begin{equation}
[\#f(0+)]= \left[\,[\#t]^{(1)}\quad [\#t]^{(2)}\,\right]\cdot
\left[\begin{array}{c} B_1\\ B_2\end{array}\right]\,,
\end{equation}
where $B_1$ and $B_2$ are unknown scalars;
the other two eigenvalues of $[\=Q]$ describe waves that amplify as $z\to\infty$
and cannot therefore contribute to the Dyakonov--Tamm wave.
At the same time,
\begin{equation}
[\#f(0-)]=\left[\begin{array}{cc}
0 &\frac{\alpha_s}{\ko}\\[5pt]
1&0\\[5pt]
\frac{\alpha_s}{\ko}\,\etao^{-1}&0\\[5pt]
0&-n_s^2\,\etao^{-1}
\end{array}\right]\cdot
\left[\begin{array}{c} A_1\\ A_2\end{array}\right]\,,
\end{equation}
by virtue of (\ref{eqn:Esub}) and (\ref{eqn:Hsub}). Continuity of the tangential
components of the electric and magnetic field phasors across the plane
$z=0$ requires that
\begin{equation}
[\#f(0-)]=[\#f(0+)]\,,
\end{equation}
which may be rearranged as
\begin{equation}
[\=M]\cdot\left[\begin{array}{c}A_1\\A_2\\B_1\\B_2\end{array}\right]=
\left[\begin{array}{c}0\\0\\0\\0\end{array}\right]\,.
\end{equation}
For a nontrivial solution, the 4$\times$4 matrix $[\=M]$ must be singular,
so that
\begin{equation}
{\rm det}\,[\=M]= 0
\label{eq:DTdisp}
\end{equation}
is the dispersion equation for the Dyakonov--Tamm wave.

\section{NUMERICAL RESULTS AND DISCUSSION}\label{res}

\begin{figure}
    \begin{center}
    \begin{tabular}{c}
    \includegraphics[height=5cm]{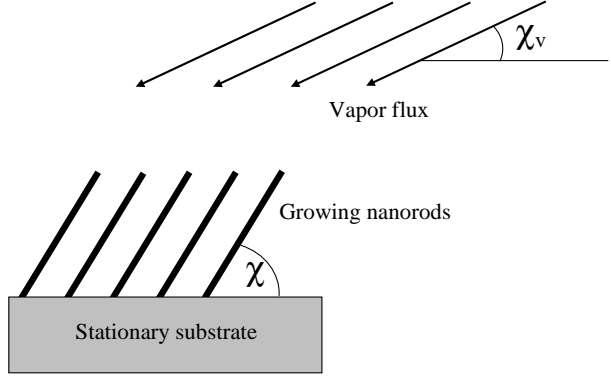}
    \end{tabular}
    \end{center}
    \caption{ \label{Fig:CTF} Schematic of the growth of a columnar thin film. The vapor flux
is directed at an angle $\chi_v$, whereas nanorods grow at an angle $\chi\geq\chi_v$. }
 \end{figure}

Although chiral STFs may be made by evaporating a wide variety of materials \cite[Chap. 1]{STFbook}, the
constitutive parameters  of chiral STFs have not been extensively measured. However, the constitutive
parameters of certain columnar thin films (CTFs) are known.  CTFs are assemblies of nanorods oriented at
an angle $\chi$ to the substrate and are produced by directing the vapor at an angle $\chi_v$ onto a
\emph{stationary} substrate, as shown in Fig.~\ref{Fig:CTF}; the vapor incidence angle
$\chi_v$ (in addition to the evaporant species)
 determines the constitutive parameters \cite{HWH1998}. When the substrate is \emph{rotated}
about a normal passing through it  at a constant angular velocity of reasonable magnitude, parallel
nanohelixes grow instead of parallel nanorods, and a chiral STF is deposited instead of a CTF
\cite{STFbook,MVS}. Although the substrate is nonstationary, the functional relationships connecting
$\eps_{a,b,c}$ and $\chi$ to $\chi_v$ for CTFs would substantially apply for chiral STFs, since the
vapor incidence angle $\chi_v$ remains constant during the deposition of thin films of either kind.

Among the CTFs which have been characterized are those made of titanium oxide, a material important in many practical
applications \cite{PNL2007}.
Empirical relationships
have been determined  for titanium--oxide CTFs at $\lambdao=633$ nm by Hodgkinson {\em et al.}
\cite{HWH1998} as
\begin{eqnarray}
&&\epsilon_a=\left[1.0443+2.7394\left(\frac{\chi_v}{\pi/2}\right)-1.3697\left(\frac{\chi_v}{\pi/2}\right)^2\right]^2\,,\label{eq:H1}\\
&&\epsilon_b=\left[1.6765+1.5649\left(\frac{\chi_v}{\pi/2}\right)-0.7825\left(\frac{\chi_v}{\pi/2}\right)^2\right]^2 \,,\label{eq:H2}\\
&&\epsilon_c=\left[1.3586+2.1109\left(\frac{\chi_v}{\pi/2}\right)-1.0554\left(\frac{\chi_v}{\pi/2}\right)^2\right]^2 \,,\label{eq:H3}
\end{eqnarray}
and
\begin{eqnarray}
\tan\chi&=&2.8818\tan\chi_v\,, \label{eq:H4}
\end{eqnarray}
where $\chi_v$ and $\chi$ are in radians.  We must caution that the foregoing expressions are applicable
to CTFs produced by one particular experimental apparatus, but may have to be modified for  CTFs
produced by others on different apparatuses; hence, we used these expressions for the numerical results
presented in this section for chiral STFs simply for illustration. Furthermore, we set $h=1$,
$\Omega=197$~nm, and $\chi_v=7.2^\circ$. Following Walker {\em et al.} \cite{Walker1998} and Polo {\em
et al.} \cite{PNL2007}, we left   $\psi$ and $n_s$ as variable parameters. All numerical results
presented in this section were computed for $\lambdao=633$~nm.

As mentioned in Section~\ref{form}, the matrix $[\=N]$ can be calculated using  two numerical techniques
\cite{PL04}: (i) the piecewise uniform approximation technique and a series technique based on the
Maclaurin expansion of $[\#P(\zeta,\kappa)]$ with respect to $\zeta$. Both yield the same results, and
the piecewise uniform approximation technique was selected for calculations reported here. Basically,
the technique consists of subdividing the chiral STF into a series of electrically thin sublayers
parallel to the interface, and assuming the dielectric properties to be spatially uniform in each
sublayer.  The accuracy of this technique depends on the thickness of the sublayers, with thinner ones
yielding more accurate results.  Based on experience \cite{STFbook,PL04}, a sublayer thickness of 2 nm
gives reasonable results.

\begin{figure}
    \begin{center}
    \begin{tabular}{c}
    \includegraphics[height=7cm]{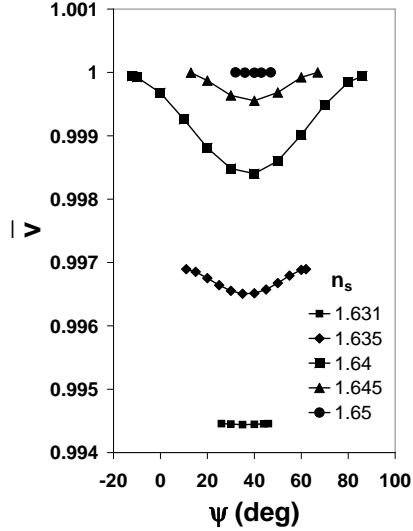}
    \end{tabular}
    \end{center}
    \caption{ \label{Fig:vbarLambda633} $\overline{v}$ as a function
    of $\psi$  with $\chi_v=7.2^\circ$ for $n_s=$ 1.631, 1.635, 1.64, 1.645, and 1.65.}
 \end{figure}

The magnitude of the phase velocity of the Dyakonov--Tamm wave was compared with that of the phase
velocity of the electromagnetic wave in the bulk isotropic material.  For this purpose, we defined the
relative phase speed
\begin{equation}
\overline{v}\equiv v_{DT}/v_s\,,
\end{equation}
where $v_{DT}=\omega/\kappa $ is the phase speed of the Dyakonov--Tamm wave and $v_s=1/n_s\sqrt{\epso\muo}$ is the phase speed of the
electromagnetic wave in the bulk isotropic material. Figure~\ref{Fig:vbarLambda633} shows $\overline{v}$
as a function of $\psi$ for several values of $n_s$. The phase velocity of the Dyakonov--Tamm wave, like
several other surface waves \cite{Dyakonov1988}--\cite{PNL2007}, was found to be lower in magnitude than
the phase velocity of the electromagnetic wave in the bulk isotropic material.

The minimum and maximum values of $n_s$  (1.631 and 1.65, respectively) in
Figure~\ref{Fig:vbarLambda633} represent the approximate limits of the $n_s$--range for which the
determinantal equation (\ref{eq:DTdisp}) representing the boundary conditions between the two material
could be solved. Outside this $n_s$--range, the Dyakonov--Tamm wave can not exist for the chosen
parameters.

Each of the curves in Figure~\ref{Fig:vbarLambda633} was drawn over the {continuous} $\psi$--range for
which the Dyakonov--Tamm wave was found to exist.  The plot in Figure~\ref{Fig:vbarLambda633} is
restricted to $\psi\in[-20^\circ,100^\circ]$.  For each value of $\psi$ in this plot (and all remaining
plots in this paper), a similar point exists at $\psi+180^\circ$ with an identical value of the
function; thus, for each value of $n_s$, there exist two separate ranges of
$\psi\in[-180^\circ,180^\circ]$ over which the Dyakonov--Tamm waves exist. Of the curves presented, the
one for $n_s=1.4$, about mid--range in $n_s$, has the widest range in $\psi$ with a width
$\Delta\psi=98^\circ$.  The mid-point of the $\psi$--range is $\psi_m=37^\circ$.  As $n_s$ approaches
either end of the $n_s$--range, $\Delta\psi$ diminishes.  At the low end of the $n_s$--range,
$\Delta\psi=51^\circ$ when $n_s=1.635$, but $\Delta\psi=20^\circ$ when $n_s=1.631$. At the high end of
the $n_s$ range, $\Delta\psi=54^\circ$ when $n_s=1.645$, but $\Delta\psi=15^\circ$ when $n_s=1.65$. Only
a slight variation in $\psi_m$, the mid--point of the $\psi$--range, is seen. For $n_s= 1.631$, 1.635,
1.4, 1.645, and 1.65, we find $\psi_m=36^\circ, 36.5^\circ, 37^\circ, 40^\circ$, and $39.5^\circ$,
respectively.  There seems to be a slight increase in $\psi_m$ as $n_s$ increases.  The values of
$\psi_m$, however, are only approximate since the end--points of the $\psi$--range were only determined
by the last whole degree lying inside the range.  This may account for why $\psi_m$ at $n_s=1.65$ is
slightly lower than that at $n_s=1.645$.

Every curve in  Figure~\ref{Fig:vbarLambda633} is smooth with a broad minimum in the vicinity of $35^\circ$ to
$40^\circ$ which levels off at both ends of the $\psi$--range. The minimum is deepest for curves
representing mid--range values of $n_s$, while curves at extreme values of $n_s$ are nearly flat. As
$n_s$ decreases, the $\overline{v}$ vs. $\psi$ curve shifts downward.

\begin{figure}
    \begin{center}
    \begin{tabular}{ccc}
    \includegraphics[height=7cm]{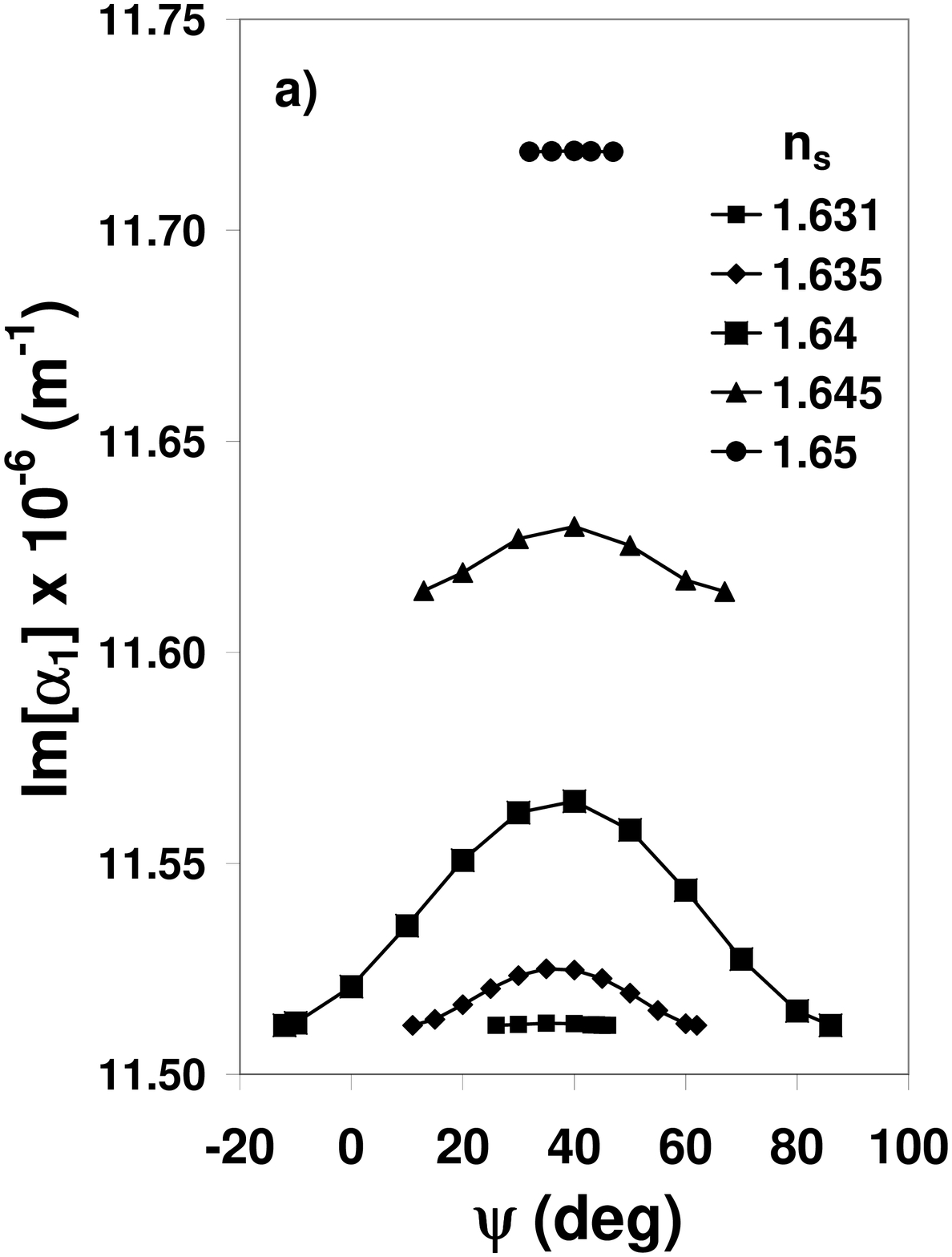}\\
    \includegraphics[height=7cm]{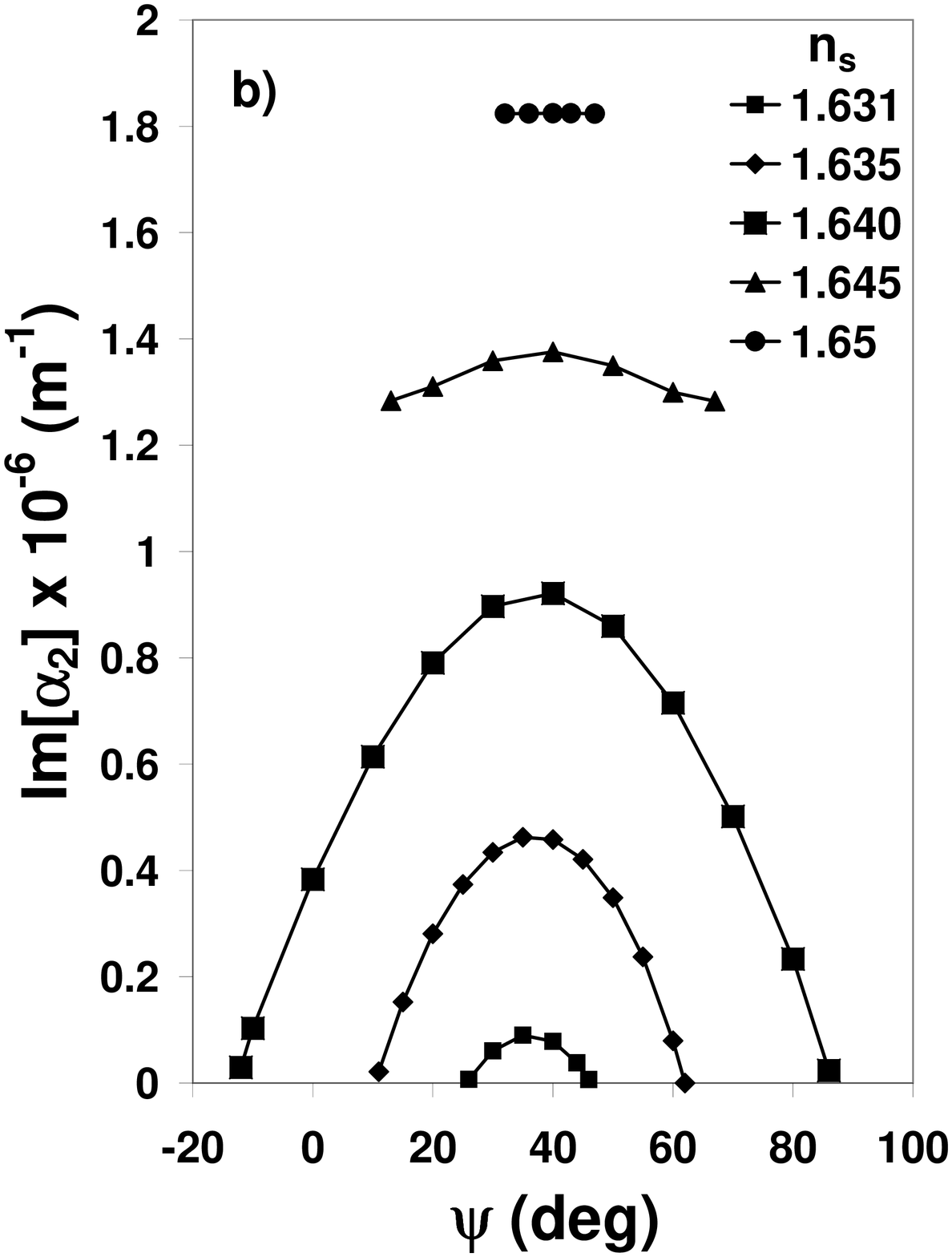}\\
    \includegraphics[height=7cm]{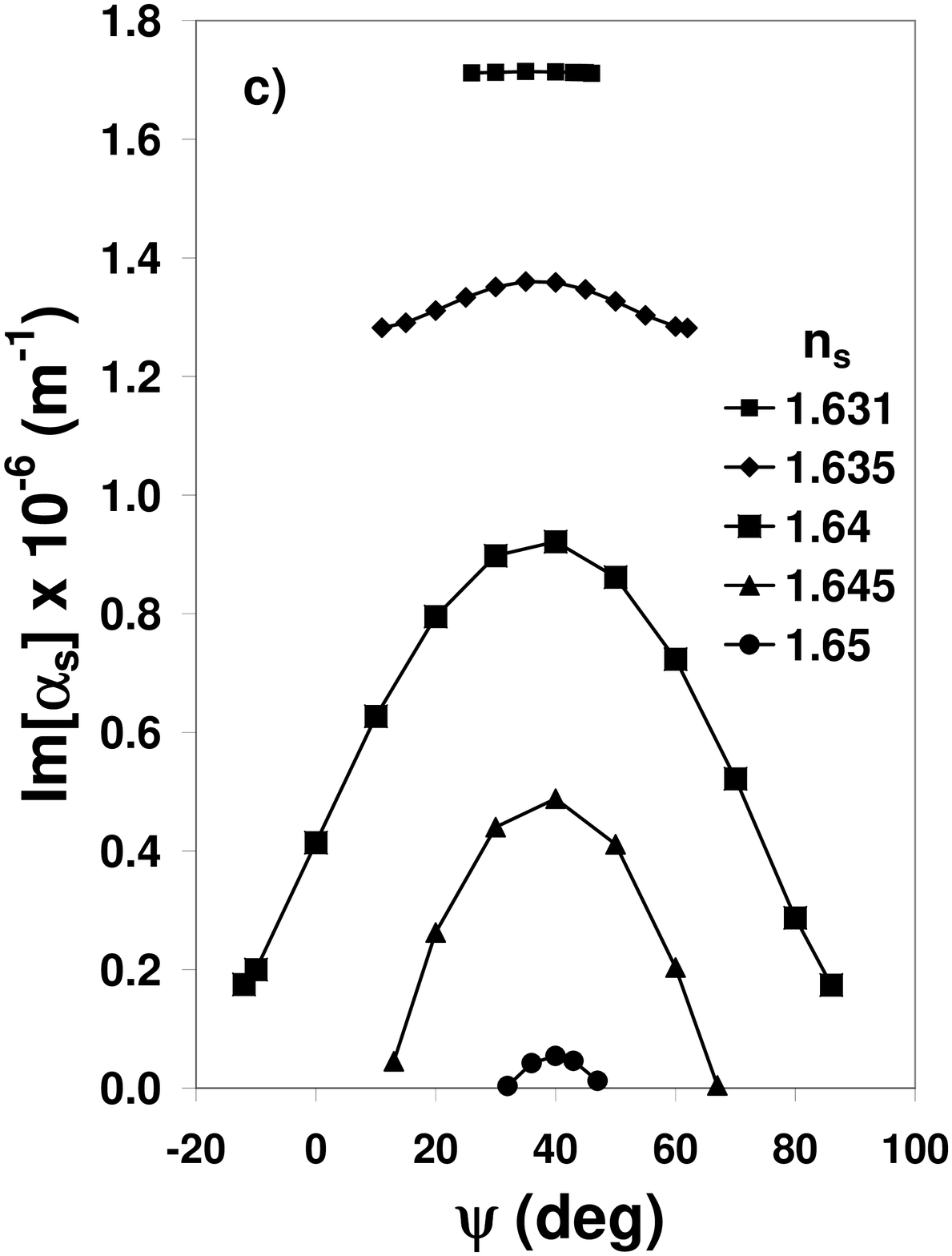}
    \end{tabular}
    \end{center}
    \caption{\label{Fig:alphas633} Decay constants  as functions of $\psi$ at $\lambdao=633$ nm for same values of $n_s$ and $\chi_v$ as in Figure
    \ref{Fig:vbarLambda633}. a) $\alpha_1$\:  b) $\alpha_2$\:  c) $\alpha_s$.}
 \end{figure}

The confinement of the Dyakonov--Tamm wave to the interface is described by the decay constants which are given by
the imaginary part of the two eigenvalues in the chiral STF ($\alpha_1$ and $\alpha_2$) and the single
eigenvalue $\alpha_s$ in the isotropic dielectric material. We found all three eigenvalues to be purely imaginary, which are shown in  Figure~\ref{Fig:alphas633} as functions
of $\psi$ for several values of $n_s$.

In Figure~\ref{Fig:alphas633}, typical values of
 ${\rm Im}[\alpha_1]$ are
about one order of magnitude larger than ${\rm Im}[\alpha_2]$.  Every  ${\rm Im}[\alpha_1]$ vs. $\psi$
curve in Figure~\ref{Fig:alphas633}a is bell--shaped with a maximum in the vicinity of $\psi=35^\circ$
to $40^\circ$.  The maximum change in ${\rm Im}[\alpha_1]$ over the $\psi$--range is greatest for
$n_s=1.64$, the mid-range of $n_s$. At this value of $n_s$, the variation in ${\rm Im}[\alpha_1]$ is
still less than 0.5\%.  Toward each end of the $n_s$--range, the ${\rm Im}[\alpha_1]$ vs. $\psi$ curve
flattens.

The   ${\rm Im}[\alpha_2]$ vs. $\psi$ curves in Figure~\ref{Fig:alphas633}b show two types of behavior.
At the high end of the   $n_s$--range ($n_s=1.645, 1.65$), the curves are bell--shaped with greater
variation occurring at $n_s=1.645$ away from the end of the $n_s$--range, just as evinced by ${\rm
Im}[\alpha_1]$.  In contrast, in the lower half of the $n_s$--range ($n_s=1.631, 1.635, 1.64$), the
${\rm Im}[\alpha_2]$ vs. $\psi$ curves do not level off at the ends of the respective $\psi$--ranges,
but continue to decrease all the way to zero.

The behavior of the decay constant in the isotropic material can be gleaned from the ${\rm
Im}[\alpha_s]$  vs. $\psi$ curves in Figure~\ref{Fig:alphas633}c.  These curves look very similar to
those of ${\rm Im}[\alpha_2]$ in Figure~\ref{Fig:alphas633}b.  However, the dependence of the shape of
the ${\rm Im}[\alpha_s]$  vs. $\psi$ curve on $n_s$ is opposite that of the ${\rm Im}[\alpha_2]$  vs.
$\psi$ curve.  The curves in Figure~\ref{Fig:alphas633}c for low values of $n_s$ are bell--shaped with
${\rm Im}[\alpha_s]$ remaining non--zero over the entire $\psi$--range; but the curves for large values
of $n_s$ maintain their downward curvature  with ${\rm Im}[\alpha_s]$ going to zero at both ends of the
respective $\psi$--ranges.   Thus, the Dyakonov--Tamm wave becomes delocalized from the interface $z=0$
at the limiting values of $n_s$; delocalization occurs on the chiral STF side of the interface at low
values of $n_s$, but on the isotropic substrate side of the interface for high values of $n_s$.

\begin{figure}
    \begin{center}
    \begin{tabular}{ccc}
    \includegraphics[height=7cm]{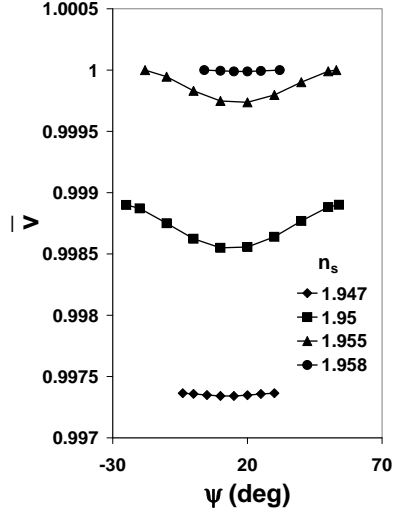}
    \end{tabular}
    \end{center}
    \caption{\label{Fig:vbarChiv25} $\overline{v}$ as a function of $\psi$ with $\chi_v=25^\circ$ for
    $n_s=$ 1.947, 1.95, 1.955, 1.958.}
 \end{figure}

Similar results are obtained at other values of $\chi_v$.  As an example, Figure \ref{Fig:vbarChiv25}
displays $\overline{v}$ when $\chi_v=25^\circ$.  As in Figure \ref{Fig:vbarLambda633}, the maximum and
minimum values of $n_s$ displayed in Figure \ref{Fig:vbarChiv25} represent the approximate limits of the
range of $n_s$ over which surface--wave propagation is possible.  The width of the $n_s$--range for
$\chi_v=25^\circ$ is roughly half of that obtained for $\chi_v=7.2^\circ$. The curves for
$\chi_v=25^\circ$ display the same general shape and trends as for the case of $\chi_v=7.2^\circ$, with
similar values of $\Delta\psi$.  However, the minimums have shifted to a lower value of $\psi$ between
$15^\circ$ and $20^\circ$.  Now, $\psi_m = 13^\circ, 14.5^\circ, 17.5^\circ,$ and $18^\circ$ for
$n_s=1.947$, 1.950, 1.955, and 1.958, respectively.  Again, a slight upward drift of $\psi_m$ as $n_s$
increases can be seen.   The figure also shows that the minimum value of $\overline{v}$ is larger for
$\chi_v=25^\circ$ than it is for $\chi_v=7.2^\circ$.  Although not shown, all three decay constants are
lower at the higher value of $\chi_v$.

Calculations were also made at $\lambdao=533$ and $733$~nm, with the assumption of the same constitutive
and geometric parameters as at $\lambdao=633$~nm. Qualitatively similar conclusions on the width of the
ranges of $n_s$ and $\psi$ for Dyakonov--Tamm waves were drawn from the numerical results obtained.
Furthermore, the minimum value of $\overline{v}$ was found to increase and the maximum values of ${\rm
Im}[\alpha_{1}]$, ${\rm Im}[\alpha_{2}]$, and ${\rm Im}[\alpha_{s}]$ were found to decrease, as the
ratio $\lambdao/\Omega$ increases.

Additional calculations, not shown here, were made for an identical chiral STF except with structural left--handedness ($h=-1$).  The exact same results were obtained with the left--handed chiral STF as with the right--handed
chiral STF presented in Figures~2--4.  It must borne in mind that, by changing $h$
from $\pm1$ to $\mp1$ in (\ref{SzDyadic}), we invert not only the structural handedness of the chiral STF but also the sense of
rotation brought about by $\psi\ne0$.

\section{CONCLUDING REMARKS}

To conclude, we examined the phenomenon of surface--wave propagation at the
planar interface of an isotropic dielectric material and a chiral sculptured thin film.
The boundary--value problem was formulated by marrying the usual
formalism for the Dyakonov wave at the
planar interface of an isotropic dielectric material and
a columnar thin film with the methodology for Tamm states in solid--state
physics. The solution of the boundary--value problem let us deduce the
existence of Dyakonov--Tamm waves.

In constitutive terms, the major difference between a CTF and a chiral STF is the
periodic nonhomogeneity enshrined in $\=S_z(z)$; in the limit $\Omega\to\infty$, a
nanohelix uncurls into a nanorod, and a chiral STF transmutes into a CTF. However,
the distinction between chiral STFs and CTFs is nontrivial, as may be deduced from
the Floquet--Lyapunov theorem \cite{YS75,LWprsa97}. In comparison to the Dyakonov wave localized to the planar interface of an isotropic dielectric material
and a CTF \cite{PNL2007}, we found that the $n_s$--range for the existence
of a Dyakonov--Tamm wave at the planar interface of an isotropic dielectric material
and a chiral STF is smaller. However, the $\psi$--range is much larger in width: in comparison to
 $\Delta\psi <1^\circ$ with  CTFs \cite{PNL2007}~---~and $\Delta\psi<
5^\circ$ with effectively uniaxial, short--period photonic crystals \cite{Artigas2005}~---~the 
width of the $\psi$--range is as high as $98^\circ$ in
Figure~\ref{Fig:vbarLambda633} with chiral STFs. This implies that
 Dyakonov--Tamm waves could be detected much more easily than Dyakonov waves.

\vspace{1 cm}
\noindent {\bf Acknowledgment.} A.L. gratefully acknowledges
a fascinating discussion with Profs. J.~Martorell (Universidad de Barcelona) and D.W.L.~Sprung (McMaster University).

\end{document}